# Comment on "Alternative interpretation of the recent experimental results of angle-resolved photoemission spectroscopy on GaMnAs [Sci. Rep. 6, 27266 (2016) ]" by M. Kobayashi *et al.*, arXiv:1608.07718

S. Souma[1], L. Chen[1], R. Oszwałdowski[2], T. Sato[3], F. Matsukura[1,4,5], T. Dietl[1,6,7], H. Ohno[1,4,5], and T. Takahashi[1,3]

[1]*WPI Research Center, Advanced Institute for Materials Research, Tohoku University, Sendai 980-8577, Japan*
[2]*Department of Physics, South Dakota School of Mines and Technology, Rapid City, SD 57701, USA*
[3]*Department of Physics, Tohoku University, Sendai 980-8578, Japan*
[4]*Center for Spintronics Integrated Systems, Tohoku University, 2-1-1 Katahira, Aoba-ku, Sendai 980-8577, Japan*
[5]*Laboratory for Nanoelectronics and Spintronics, Research Institute of Electrical Communication, Tohoku University, 2-1-1 Katahira, Aoba-ku, Sendai 980-8577, Japan*
[6]*Institute of Physics, Polish Academy of Sciences, aleja Lotników 32/46, PL-02-668 Warszawa, Poland*
[7]*Institute of Theoretical Physics, Faculty of Physics, University of Warsaw, ulica Pasteura 5, PL-02-093 Warszawa, Poland*

**Abstract**

Recently, Kobayashi *et al.* (arXiv:1608.07718; ref. 1) have proposed an alternative interpretation of our angle-resolved photoemission spectroscopy (ARPES) results for the dilute ferromagnetic semiconductor (Ga,Mn)As. They claim that our ARPES data [Sci. Rep. **6**, 27266 (2016); ref. 2] can be explained by locating the Fermi level $E_F$ above the valence-band top, supporting the impurity-band model of ferromagnetic (Ga,Mn)As. In this comment, we show that the assignment of bands' positions in respect to $E_F$ by Kobayashi *et al.* is not consistent with our data. By comparing the ARPES result to band-structure calculations, we demonstrate clearly that $E_F$ resides inside the valence band in accordance with the *p-d* Zener model.

In our ARPES study of (Ga,Mn)As [2], we have found that the Fermi level $E_F$ is located inside the valence band (VB) of host GaAs, which supports the *p-d* Zener model of the carrier-mediated ferromagnetism in (Ga,Mn)As [3,4]. This model, the so-called VB model, is also corroborated by other experimental results and functionalities such as the carrier-doping dependence of Currie temperature $T_C$, the strain or hole-concentration dependence of magnetic anisotropy, the magnitude of carrier spin polarization and the tunnelling magnetoresistance [5]. The latter paper [5] contains also a comprehensive discussion on how to interpret, often quantitatively, those findings that were initially considered, in a qualitative manner, as supporting the impurity band conjecture. It is worth adding that the VB model is substantiated by the state-of-the-art *ab initio* results [6].

Recently, Kobayashi *et al*. have argued [1] that our ARPES data may involve ambiguity in estimating the energy position of the VB, and they proposed an alternative interpretation, namely that $E_F$ is located above the VB top, which would support the impurity-band model. Here we show that this alternative interpretation is not consistent with our ARPES data.

Kobayashi *et al*. insist that the experimental band structure in the wide energy region ($E_F$ - 4 eV) is well reproduced with the tight-binding calculation of GaAs when the VB top is situated at 40 meV below $E_F$ [1]. In Fig. 1 we show our experimental results for samples with different $T_C$ values (Fig. 2g in ref. 2) compared to the outcome of the tight-binding computations of the GaAs band structure assuming that the Fermi energy is above the VB top, as claimed by Kobayashi *et al*. As seen, there is a striking disagreement between the experiment and the calculation, in particular in the region near the top of the split-off (SO) valence band. Actually, a simple rigid-band shift of the calculated bands of host GaAs is not a good approximation to determine the accurate $E_F$ position in (Ga,Mn)As. Considering this fact, in our paper we identified the bands' origin by comparing only their dispersion with the calculations (Fig. 1 in ref. 2). This method of the band assignment reveals that the steep band dispersion reaching $E_F$ (band B in Fig. 1 in ref. 2) should be assigned to the SO subband. Importantly, this assignment is corroborated by Kobayashi *et al*. [1] and Kanski *et al*. [7]. It is also noted that while Kobayashi *et al*. cautioned that the momentum distribution curve (MDC) analysis is not reliable at the top/bottom of bands, we had already taken into account this point and applied also the energy distribution curve (EDC) analysis in determining the VB dispersion.

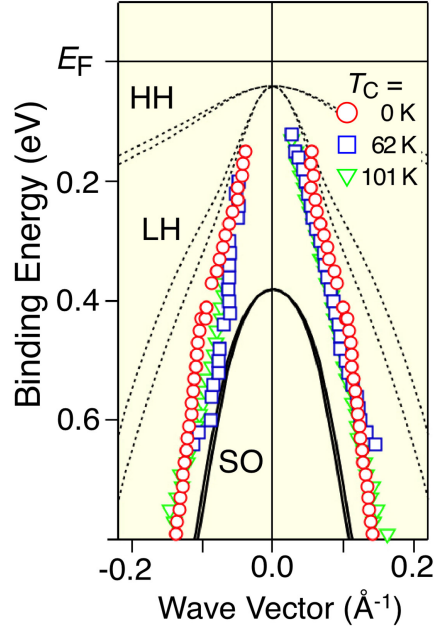

Fig. 1: Plot of band dispersions of (Ga,Mn)As with $T_C$ = 101 K and 62 K and n-GaAs, extracted from the peak position of momentum distribution curves (MDCs). Experimental data are the same as Fig. 2g in ref. 2. Solid and dotted black curves show the results of band-structure calculations for GaAs shifted downward to match the $E_F$ position with ref. 1. Note that the top of the SO band in the experiment is not reproduced by the calculation.

In ref. 2, we have demonstrated that the intensity of the SO band appears to reach $E_F$, accompanied with a Fermi-edge cut-off, indicating its metallic nature. Since the spectral intensity around the SO-band top is strongly suppressed by disorder and/or matrix-element effects [2], it was difficult to estimate the accurate location of the band top from the peak position of EDCs. Nevertheless, since the left-hand branch of the SO band in Figs. 2d and 2f in ref. 2 is visible at 0.1 eV and is separated from the right-hand counterpart, one can conclude that the top of the SO band is very close to $E_F$ (note that in ref. 2, we did not plot the experimental band dispersion around the top of the SO band to avoid possible error in the MDC analysis, as pointed out in ref. 1). By taking into account the energy difference of ~ 0.35 eV between $\Gamma_7$ and $\Gamma_8$ states (this value would not markedly change upon Mn doping [3]), it is quite likely that the top of the heavy/light hole (HH/LH) subbands is located at 0.25 eV above $E_F$. Thus, our ARPES data strongly suggest *without fitting to any theoretical curve* that $E_F$ of (Ga,Mn)As is deep inside the VB.

Kobayashi *et al*. have suggested that the observed SO band is an "unknown surface state". As seen in Fig. 2 of ref. 2, the SO band displays the Λ-shaped feature, whereas that of GaAs is rounded (parabolic-like). If the Λ-shaped feature arises from the surface states, it should be sensitive to surface aging. However, we have not observed its particular sensitivity to the surface aging in comparison with bulk bands. This suggests the bulk origin of the SO band, consistent with a recent ARPES study by Kanski *et al*.

[7]. Our ARPES data suggest that the variation of the SO-band dispersion from parabolic to Λ-shape upon Mn doping reflects an intrinsic change of electronic states, but not the surface effect or some other extrinsic effects. It is known that the Mn-impurity level located at 110 meV above the VB top gradually approaches the VB top upon Mn doping [7,8]. This suggests that the variation of the SO-band dispersion may be related to the enhanced interactions between the VB and the Mn-derived states upon Mn doping. It is also suggested that the variation of the SO-band dispersion can be related to the disorder-induced correlation effect, since the soft Coulomb gap has been observed near $E_F$ [2].

Next we discuss the exchange splitting of the SO band. From the tight-binding and $kp$ calculations, the exchange splitting of the SO band for the $T_C \sim 100$ K sample is ~ 60 meV at 30 K [3]. Such a splitting has not been clearly resolved in ref. 2, likely due to strong broadening of ARPES spectrum by Mn-induced scattering. In fact, as seen in Fig. 4a of ref. 2, the MDC width in the experiment is much broader than that expected from the calculated bands incorporating the exchange splitting. Considering such an intrinsic spectral broadening, we carried out the magnetic linear dichroism (MLD) measurement [2]. A MLD signal is not observed in nonmagnetic pristine GaAs. In contrast, a signature of MLD in the SO band is clearly seen as asymmetric intensity pattern only below $T_C$ in (GaMn)As. Numerical simulation shows that the exchange splitting of several tens of meV is necessary to reproduce the MLD effect in the SO band [2]; this value is comparable to that in our ARPES experiment. Our results thus strongly support the existence of the expected large spin-splitting due to $p$-$d$ hybridization [3] and contradict the impurity-band model which predicts almost negligible exchange splitting (< 3 meV) [1].

Finally, we comment on the invariance of the SO band between $T_C$ = 101 and 62 K seen in Fig. 2 of ref. 2, which might be inconsistent with the change in the carrier density expected from $T_C$. According to ref. 4, if there is no difference in the Mn content, so that the lowering of $T_C$ results entirely from a decrease in the hole concentration, the position of $E_F$ would differ by 30 meV at 30 K (see, Fig. 1a in ref. 2 for temperature dependence of magnetization). As described above, there exists a strong spectral broadening in the ARPES spectrum. This broadening, together with a dependence of the Coulomb gap on $T_C$ (Fig. 3 in ref. 2), causes the uncertainty of 20-30 meV in determining the band location near $E_F$. If the energy shift of the SO band is comparable to this value, it would not be clearly resolved in the ARPES experiment. To establish the overall energy shift of the VB, it is important to experimentally clarify the location of the HH/LH bands which produce much larger Fermi surfaces. In this regard, it would be useful to perform ARPES experiment with synchrotron light to enhance the cross section of the HH/LH bands.

In summary, we have shown that the Fermi level is located inside the VB of (Ga,Mn)As, contrary to the suggestion of Kobayashi *et al*. [1]. Our result strongly suggests that the VB hole carriers mediate the ferromagnetic interaction among localized Mn moments in accordance with the *p-d* Zener model.

**REFERENCES**
[1] M. Kobayashi *et al*., arXiv:1608.07718.
[2] S. Souma *et al*., Sci. Rep. **6**, 27266 (2016).
[3] T. Dietl, H. Ohno, F. Matsukura, J. Cibert, and D. Ferrand, Science **287**, 1019 (2000).
[4] T. Dietl, H. Ohno, and F. Matsukura, Phys. Rev. B **63**, 195205 (2001).
[5] T. Dietl and H. Ohno, Rev. Mod. Phys. **86**, 187 (2014).
[6] I. Di Marco *et al.,* Nat. Commun. **4**, 2645 (2013); K. Sato *et al.,* Rev. Mod. Phys. **82**, 1633 (2010).
[7] J. Kanski *et al*., arXiv:1608.06821.
[8] T. Jungwirth *et al*., Phys. Rev. B **76**, 125206 (2007).